\documentclass[12pt]{article}

\usepackage{newtxtext,newtxmath}

\usepackage{graphicx}

\usepackage[letterpaper,margin=1in]{geometry}

\renewenvironment{abstract}
	{\quotation}
	{\endquotation}

\date{}

\makeatletter
\renewcommand{\fnum@figure}{\textbf{Figure \thefigure}}
\renewcommand{\fnum@table}{\textbf{Table \thetable}}
\makeatother

\usepackage{scicite}
\usepackage{float}

\usepackage{url}

\def\scititle{
	Recurrent neural networks implemented through spatiotemporal light propagation in optical fibers
}
\title{\bfseries \boldmath \scititle}

\author{
	Dilem E\c{s}lik$^{1,\dagger}$,
	Bahad{\i}r Utku Kesgin$^{1,\dagger}$,
	U\u{g}ur Te\u{g}in$^{1,\ast}$\and
	\small$^{1}$Department of Electrical and Electronics Engineering, Ko\c{c} University, \.{I}stanbul, 34450, T\"{u}rkiye.\and
	\small$^\ast$Corresponding author. Email: utegin@ku.edu.tr\and
	\small$^\dagger$These authors contributed equally to this work.
}

\begin{document} 

\maketitle

\begin{abstract} \bfseries \boldmath
Recurrent neural networks excel at temporal tasks and video processing but require energy-intensive sequential memory operations. We demonstrate that multimode optical fibers naturally implement spatiotemporal recurrent computation through passive light propagation. Video frames are encoded onto separate optical beams with controlled time delays; these beams combine and recirculate through a fiber loop where interference and nonlinear propagation generate high-dimensional states encoding both current inputs and fading memory. Remarkably, the entire optical system remains fixed with no trainable parameters or electronic feedback, yet this single physical configuration achieves competitive performance across diverse temporal and spatiotemporal learning tasks: chaotic time-series forecasting, human action recognition, steering angle prediction, and surgical skill assessment. Our results show that recurrent temporal processing can emerge directly from spatiotemporal wave dynamics. This paradigm shift from algorithmic to physical recurrence offers an energy-efficient pathway to temporal artificial intelligence by leveraging intrinsic spatiotemporal optical nonlinearities within multimode fibers.
\end{abstract}

\noindent
\section{Introduction}
Information in physical systems often unfolds over time rather than appearing instantaneously. From predicting future events to recognizing dynamic patterns, the ability to integrate information across time enables adaptive behavior in both biological and artificial systems. Recurrent neural networks (RNNs) process temporal variations by maintaining internal states that encode memory of past inputs, allowing current outputs to depend on temporal context \cite{hochreiter1997long,pascanu2013difficulty}. However, implementing RNNs in conventional electronic hardware incurs substantial computational overhead, especially for video processing. Each recurrent update requires reading stored states from memory, computing nonlinear transformations, and writing updated states, a cycle that repeats at every time step, consuming energy proportional to both state dimensionality and sequence length \cite{pascanu2013difficulty}. Beyond machine learning, efficient temporal processing is critical for edge computing, autonomous systems, and real-time sensor analysis---applications where power budgets and processing latency are severe constraints.

Photonic computing has emerged as an alternative paradigm that exploits the intrinsic parallelism and low-latency propagation of light to accelerate machine learning operations \cite{Wetzstein2020DeepOptics,McMahon2023PhysicsOC}. Optical neural networks can perform matrix-vector multiplications through wave interference, enabling energy-efficient inference for feedforward architectures \cite{feldmann2021parallel,Lin2018Diffractive}. Programmable photonic circuits and optical memory technologies further expand the capabilities of optical processors \cite{rios2015integrated,bogaerts2020programmable}. However, extending photonic systems to recurrent temporal processing, where memory and feedback are essential, remains challenging. Most implementations require active electronic control to close recurrent loops, reintroducing the energy costs and timing constraints that photonics aims to avoid. Early demonstrations of photonic information processing highlighted the potential for ultrafast and energy-efficient computation using transient optical dynamics and nanophotonic circuits \cite{brunner2013parallel,miller2017attojoule}.

Reservoir computing offers a framework for photonic temporal processing. Reservoir computing provides an alternative training paradigm in which only a linear readout layer is optimized while the recurrent dynamics remain fixed \cite{lukovsevivcius2009reservoir}. 
In this approach, temporal signals are fed into a fixed nonlinear system, the `reservoir', whose complex internal dynamics naturally create diverse response patterns. These patterns serve as features for simple electronic classifiers, avoiding the need to train the recurrent dynamics themselves \cite{Maass2002LSM,Jaeger2004Harnessing}. Photonic delay-based reservoirs have demonstrated efficient processing of scalar time series by circulating signals through fiber loops with temporal multiplexing \cite{Appeltant2011Reservoir,Larger2017PhotonicReservoir,Bueno2018Reinforcement}. Speckle-based and multimode waveguide photonic reservoirs further highlight the potential of complex optical scattering for temporal information processing \cite{Paudel2020Speckle,Ashner2021Photonic}. These concepts have been realized across optoelectronic, passive cavity, silicon photonic and wavelength-multiplexed platforms \cite{Vandoorne2014SiliconRC,Vinckier2015PassiveRC,Martinenghi2012Transient,Larger2012LaserRC,VanDerSande2017}. However, these implementations process one-dimensional signals sequentially, limiting throughput when handling spatially structured data such as images or video frames. Extending reservoir approaches to native spatiotemporal processing, where both spatial structure and temporal dynamics are preserved throughout the optical propagation, would enable direct video analysis and broaden the applicability of photonic temporal processors. 

Multimode optical fibers offer a platform for complex spatiotemporal processing. When light propagates through these fibers, different spatial patterns (modes) interfere and interact nonlinearly, creating rich dynamics that have been exploited for imaging \cite{popoff2010measuring,vcivzmar2012exploiting} and machine learning \cite{borhani2018learning,Tegin2021SOLO}. Recent work has shown that these fibers naturally perform high-dimensional nonlinear transformations on spatial patterns \cite{wright2015controllable,wright2022nonlinear,Tegin2019SelfSimilar}, suggesting they could provide the complex dynamics needed for recurrent processing if appropriately configured with feedback.

Here we demonstrate that passive recurrence in multimode optical fibers implements recurrent temporal processing for spatiotemporal data processing. By encoding sequential video frames onto spatially multiplexed optical channels with fiber-induced delays and coupling the output back into the system through a passive fiber loop, we create a physical recurrent operator in which memory emerges from delayed interference and nonlinearity arises from Kerr-type light-matter interactions \cite{Agrawal2020NFO}. Critically, no parameters within the optical network are trained or adjusted; learning occurs solely through a linear electronic readout applied to the output optical states. By eliminating electronic state management and leveraging free optical propagation for recurrent updates, this approach reduces electronic computation and memory overhead compared to conventional implementations. This decoupling of recurrent dynamics from task-specific learning enables a single fixed optical system to process diverse temporal learning tasks.

We experimentally validate this approach across four diverse domains requiring different types of temporal reasoning: chaotic time-series prediction (long-term dependencies), human action recognition (spatiotemporal motion patterns), autonomous driving scene analysis (continuous control from visual flow), and surgical skill assessment (subtle motion quality). Across these benchmarks, the system produces task-selective optical embeddings and achieves competitive performance using only passive photonic recurrence and simple linear regression. These results demonstrate that recurrent computation, traditionally implemented algorithmically, can be realized directly through the spatiotemporal physics of light propagation in multimode fibers. We first describe the physical mechanism of passive recurrence, then evaluate performance across four application domains, and finally discuss implications for physical computing.
\section{Results}

We first express the physical mechanism of temporal processing in the optical system, then demonstrate its performance across diverse application domains.

\subsection*{Physical mechanism of recurrent optical processing}

\begin{figure}[H]
\centering
\includegraphics[width=\textwidth]{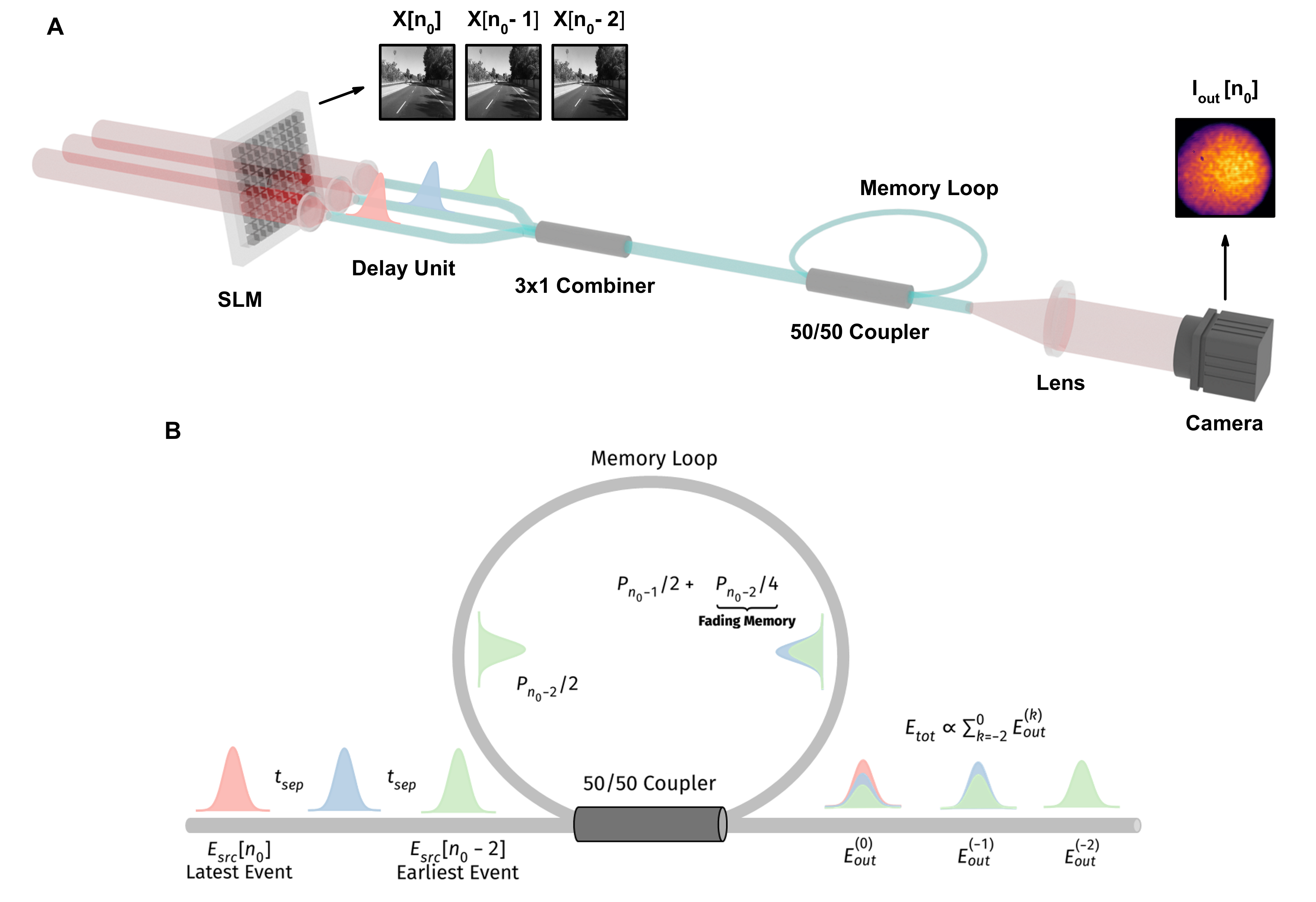}
\caption{\textbf{Experimental setup and physical mechanism of recurrent optical processing.} (\textbf{A}) Experimental schematic showing the passive recurrent architecture. Sequential video frames are spatially multiplexed onto a spatial light modulator (SLM), coupled into multimode fibers with different delay lengths, combined, and fed into a 50/50 fiber coupler loop. Half the signal recirculates while half is detected as speckle patterns. (\textbf{B}) Schematic illustrating the pulse propagation and memory mechanism within the memory loop and processing fiber after the delay unit.}
\label{fig:mechanism}
\end{figure}

The experimental architecture illustrated in Figure \ref{fig:mechanism} implements recurrence through passive optical propagation without electronic feedback or trainable parameters. Sequential video frames at discrete times $n_0$, $n_0-1$, and $n_0-2$ are encoded as phase masks onto three spatially separated regions of a spatial light modulator (SLM). Each encoded beam is coupled into a multimode fiber of different length, creating controlled time delays, before the three channels are combined using a passive fiber combiner. The combined optical field enters a 50/50 fiber coupler that splits the signal: half recirculates through a fiber loop while half is directed to a camera for readout as speckle patterns. This architecture creates natural temporal memory through two mechanisms. First, the fiber delays align temporally ordered inputs (frames at $n_0$, $n_0-1$, $n_0-2$) so they arrive sequentially at the combiner, effectively creating a temporal window. Second, the recurrent loop allows signals to circulate multiple times with progressive attenuation from the 50/50 splitting, implementing fading memory where recent inputs dominate but earlier time steps contribute with exponentially decaying influence. Here, recurrence emerges purely from passive fiber propagation and interference.

The multimode fiber propagation adds further complexity through nonlinear spatiotemporal mixing. The 200/220~$\mu$m multimode fiber supports on the order of $10^{4}$ spatial modes at 1030~nm, enabling high-dimensional spatial mixing of the encoded wavefronts. As light propagates, thousands of spatial modes interfere and interact through Kerr nonlinearity, generating high-dimensional transformations that depend on both the spatial pattern and temporal pulse structure.

Propagation of light in a multimode optical fiber is governed by a highly complex partial differential equation. Since our physical model depends on meters of propagation within the fiber of multiple beams, to express the operations in our setup analytically we define a compact function which defines a nonlinear mapping from the input field to the output field as a function of fiber length $\ell_{fiber}$:

\begin{equation}
E_{\mathrm{out}} = \mathcal{N}_{\mathrm{fiber}}(E_{\mathrm{in}};\ell_{\mathrm{fiber}}).
\end{equation}

We examine the partial differential equation governing this function and its nonlinear terms in detail in the Materials and Methods section. Within the delay unit each spatially multiplexed timestep goes through the following process:
\begin{equation}
\begin{aligned}
E_{delay \,|\, n_0} \;&=\; \mathcal{N}_{fiber}(\mathcal{F}[A_0 e^{i\phi_{X[n_0]}}]; \ell_{fixed} + 2\ell_{loop}) \\
E_{delay \,|\, n_0-1} \;&=\; \mathcal{N}_{fiber}(\mathcal{F}[A_0 e^{i\phi_{X[n_0-1]}}]; \ell_{fixed} + \ell_{loop}) \\
E_{delay \,|\, n_0-2} \;&=\; \mathcal{N}_{fiber}(\mathcal{F}[A_0 e^{i\phi_{X[n_0-2]}}]; \ell_{fixed})
\end{aligned}
\end{equation}

Where $A_0$ is the amplitude of the modulated Gaussian beam, $\mathcal{F}$ denotes 2D Fourier Transform, and $\phi_{[n_0]}$ is the phase pattern of the time step $n_0$. The difference in fiber propagations result in a temporal delay between each spatially multiplexed timestep. The beams are processed individually in the delay unit with the effect of self-phase modulation (SPM). After initiating delay,  we use a 50/50 coupler to combine separately processed timesteps in the delay unit and harness cross-phase modulation (XPM) alongside SPM to extract worthy temporal information. For each beam coming into the coupler, half of it proceeds to the processing fiber and the other half of it proceeds to the memory loop. Memory loop delays the seperated beam such that when the new timestep arrives, previous timesteps can copropagate and coprocess more recent data with the information they contain from the past. For the $k$th time step coming into the coupler, we can write:
\begin{equation}
E_{in}^{(k)} = \mathcal{N}_{fiber}(E_{delay \,|\, k};\ell_{fixed}) + \mathcal{N}_{fiber}(\frac{1}{\sqrt{2}}E_{in}^{(k-1)}; \ell_{loop})
\end{equation}
Due to our recurrence structure, $E_{in}^{(n_0 + k)} = 0$ for $k \le -3$, and $E_{delay\,|\,n_0 + k} = 0$ for $k \ge 1$. Fully expressing this recursive expression for the most recent time step ($n_0$) we obtain the following expression which analytically demonstrates the presence of a fading memory within the system:
\begin{equation}
\begin{aligned}
E_{in}^{(n_0)} =
\mathcal{N}_{fiber}\Big[
&\overbrace{\tfrac{1}{\sqrt{2}}\mathcal{N}_{fiber}(E_{delay\,|\,n_0-1};\ell_{coupler})}^{\text{Power}\,\propto P_0/2}
\\
&\quad+
\overbrace{\tfrac{1}{\sqrt{2}}\mathcal{N}_{fiber}\!\left(
\tfrac{1}{\sqrt{2}}\mathcal{N}_{fiber}(E_{delay\,|\,n_0-2};\ell_{coupler});
\ell_{loop}\right)}^{\text{Power}\,\propto P_0/4};
\ell_{loop}
\Big] +\mathcal{N}_{fiber}(E_{delay\,|\,n_0};\ell_{coupler})
\end{aligned}
\end{equation}

In these expressions $\ell_{fixed} = \ell_{coupler} = \text{1 m and}\, \ell_{loop} = \text{5 m}$. As demonstrated in the expression for the input electric field in the final timestep, the final timestep is coprocessed by two fading memory components that are the previous timesteps.  Each time step fades in terms of power by a factor of two. We capture the output of the optical recurrent neural network through a camera, since the events are happening too fast compared to camera shutter, in the output intensity we see the incoherent sum of the output electric fields from the processing port of the combiner. Since our source has a repetition rate of 9 kHz, calculating for the delay period of $24ns$, the most recent sample will fade within the memory such that when the earliest event arrives at the delay loop again, the contribution from the previous repetition is negligible. But even though there will be no incoming information, an important proportion of the most recent beam and some information from the previous events will continue circulating together until fading. Considering these, for processed timestep $n_0$ we can write the output intensity of the optical recurrent neural network as:

\begin{equation}
I_{out}[n_0] = \sum_{k\,=\,-2}^{2} |\mathcal{N}_{fiber}(E_{in}^{(n_0 + k)};\ell_{coupler})|^2
\end{equation}

Here it is important to revisit the following expressions $E_{in}^{(n_0 + k)} = 0$ for $k \le -3$, and $E_{delay\,|\,n_0 + k} = 0$ for $k \ge 1$. 

\subsection*{Chaotic time-series prediction and autonomous forecasting}

We first evaluate the optical recurrent processor on chaotic time-series prediction, a canonical test of nonlinear memory and temporal learning capability. We used the Santa Fe chaotic laser time-series benchmark \cite{hubner_dimensions_1989,noauthor_mcnames_nodate}, which tests whether the system can capture delayed nonlinear dependencies and sustain accurate long-horizon forecasting. The scalar time series is encoded onto the SLM using triplets of consecutive samples. As these temporally ordered inputs propagate through the multimode fibers and recirculate through the loop, the resulting speckle patterns form high-dimensional optical embeddings that encode both current values and delayed information.

\begin{figure}[H]
\centering
\includegraphics[width=0.5\textwidth]{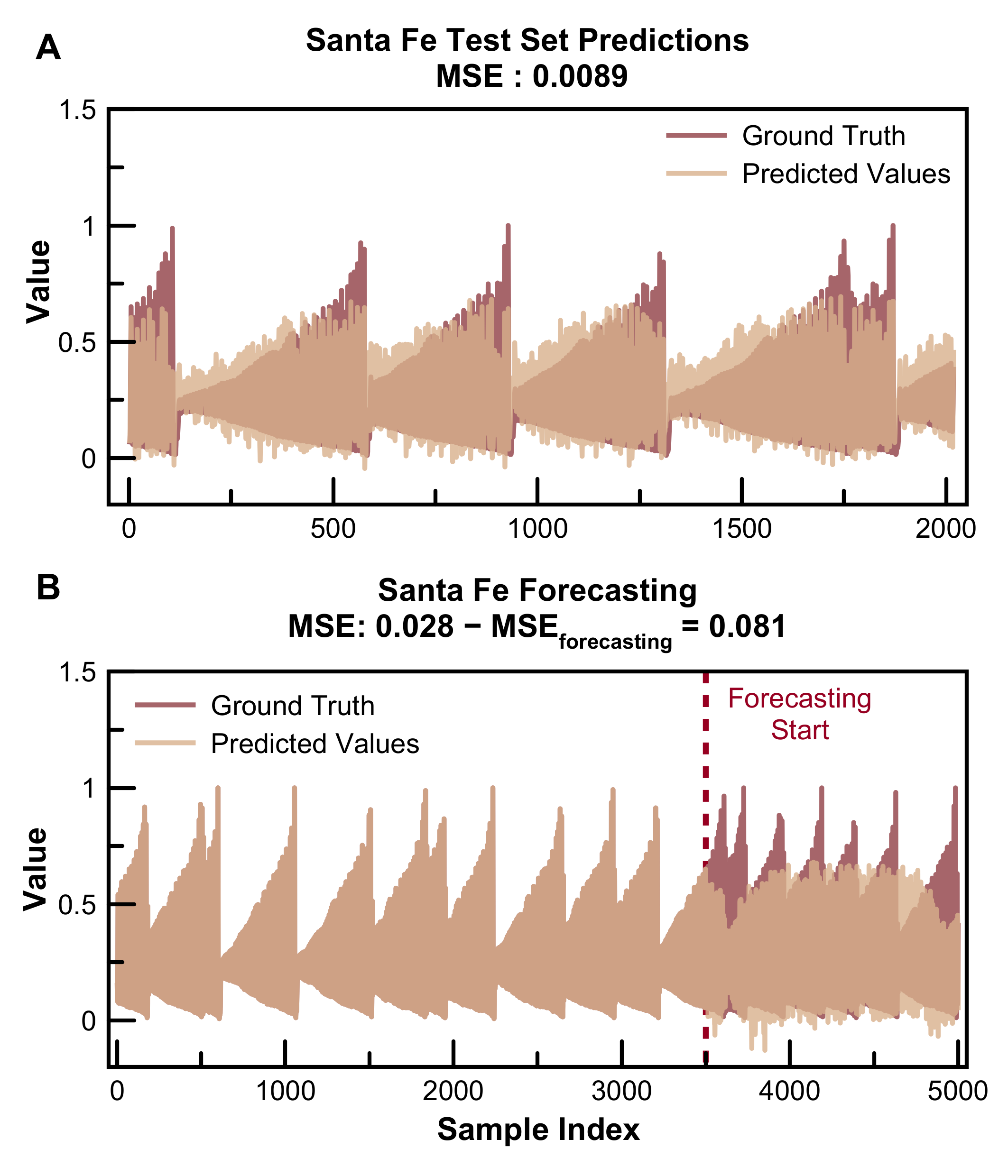}
\caption{\textbf{Chaotic time-series processing.} (\textbf{A}) One-step-ahead prediction test set results on the Santa Fe laser dataset. (\textbf{B}) Autonomous forecasting results where the system recursively predicts future states without access to ground truth.}
\label{fig:santafe}
\end{figure}

As illustrated in Figure \ref{fig:santafe}, a ridge regression readout trained on these embeddings achieve  mean squared error (MSE) of 0.084 on the test set, indicating accurate short-term prediction of the nonlinear dynamics. To assess long-horizon forecasting, we switch to an autonomous mode where predicted values were recursively fed back as inputs. In this closed-loop configuration, the system generates predictions without access to ground truth, testing its ability to capture the underlying attractor structure. The model is first trained and then autonomous forecast maintains accurate trajectory tracking for over many time steps, as shown in Figure \ref{fig:santafe}. This performance demonstrates that the passive optical recurrence captures sufficient delayed nonlinear information to sustain long-term dynamical predictions.

\subsection*{Human action recognition from video}

\begin{figure}[H]
\centering
\includegraphics[width=\textwidth]{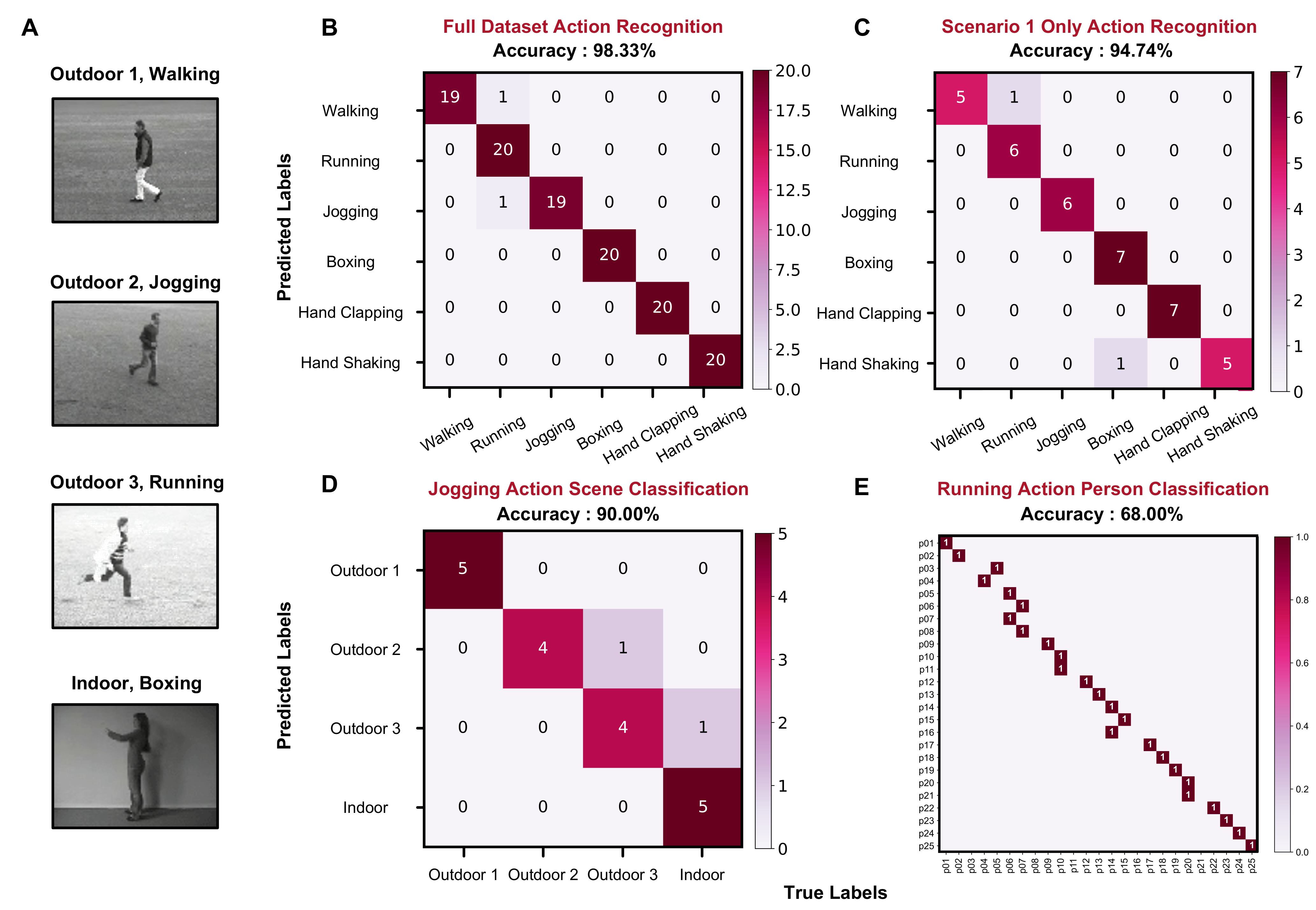}
\caption{\textbf{Human action recognition from video.} (\textbf{A}) Representative frames from the KTH dataset showing six action categories. (\textbf{B}) Confusion matrix of full-dataset action classification achieving 96.67\% accuracy. (\textbf{C}) Confusion matrix of scene-specific classification where only action varies. (\textbf{D}) Confusion matrix of scene classification using the jogging action. (\textbf{E}) Confusion matrix of person identification within the running action category.}
\label{fig:kth}
\end{figure}

Moving from scalar time series to video data, we evaluate the system on human action recognition. This is a task critical for surveillance, human-computer interaction, and activity monitoring applications. The KTH human action dataset \cite{schuldt_recognizing_2004} contains videos of six action categories (walking, jogging, running, boxing, hand-waving, hand-clapping) performd by 25 subjects across four recording environments, providing a controlled benchmark for assessing whether the optical system can extract motion-selective features from spatiotemporal video without prior training of the optical dynamics. Each video was processed by encoding consecutive frame triplets onto the SLM. The accumulated speckle pattern over all triplets in a video serve as a single compact representation, which is used as input to a ridge regression classifier. Figure \ref{fig:kth} illustrates the results of machine learning tasks on KTH dataset. On the full dataset, the optical reservoir achieves 98.33\% classification accuracy, on par with established video recognition methods despite using only passive optical processing and linear readout.

To probe which features the optical system encodes, we perform selective classification experiments on the KTH dataset. For general action classification and action-specific scene classification, an 80/20 train–test split was used, while a 75/25 split was used for Scenario-1 action classification and person identification due to the limited number of samples per class. When the scene was fixed and only the action varied, classification accuracy reached 94.74\% , indicating robust action-selective embeddings. Conversely, when the action was fixed and the scene varied, the same optical representations enabled scene classification . Finally, fixing the action allowed subject identification across 25 individuals . These results demonstrate that the optical states simultaneously encode multiple aspects of spatiotemporal structure—motion dynamics, contextual background, and identity-specific variations—without requiring separate processing pathways or task-specific optical configurations. For comparison with prior work, Table \ref{tab:kth} reports action-recognition results obtained under the standard Scenario-1 protocol commonly used in the literature. The passive optical recurrent processor achieves performance comparable to convolutional-recurrent networks and photonic reservoir approaches while using a single fixed optical architecture and minimal electronic processing.

\begin{table}[H]
    \centering
    \caption{\textbf{Performance comparison on KTH human action recognition.} Fixed scene refers to scene-specific action classification for the first scene. Train/test splits vary across methods as reported in original publications.}
    \begin{tabular}{l c c c}
        \hline
        \textbf{Method} & \textbf{Train/Test Split} & \textbf{Fixed Scene} & \textbf{Full Dataset} \\
        \hline
        Interest Points + SVM \cite{yadav_action_2016} & 80\% - 20\% & - & 98.20\% \\
        \textbf{This Work (Optical RNN)} & \textbf{75\% - 25\% / 80\% - 20\%} & \textbf{94.74\%} & \textbf{98.33\%} \\
        CNN \& RNN \cite{baccouche_sequential_2011}  & 64\% - 36\% & - & 94.39\% \\
        Differential RNN \cite{veeriah_differential_2015} & 64\% - 36\% & - & 93.96\% \\
        Photonic RC \cite{antonik_human_2019} & 75\% - 25\% & 91.3\% & - \\
        LSTM \cite{grushin_robust_2013} & 64\% - 36\% & - & 90.7\% \\
        \hline
    \end{tabular}
    \label{tab:kth}
\end{table}

\subsection*{Steering angle prediction for autonomous driving}

\begin{figure}[H]
\centering
\includegraphics[width=\textwidth]{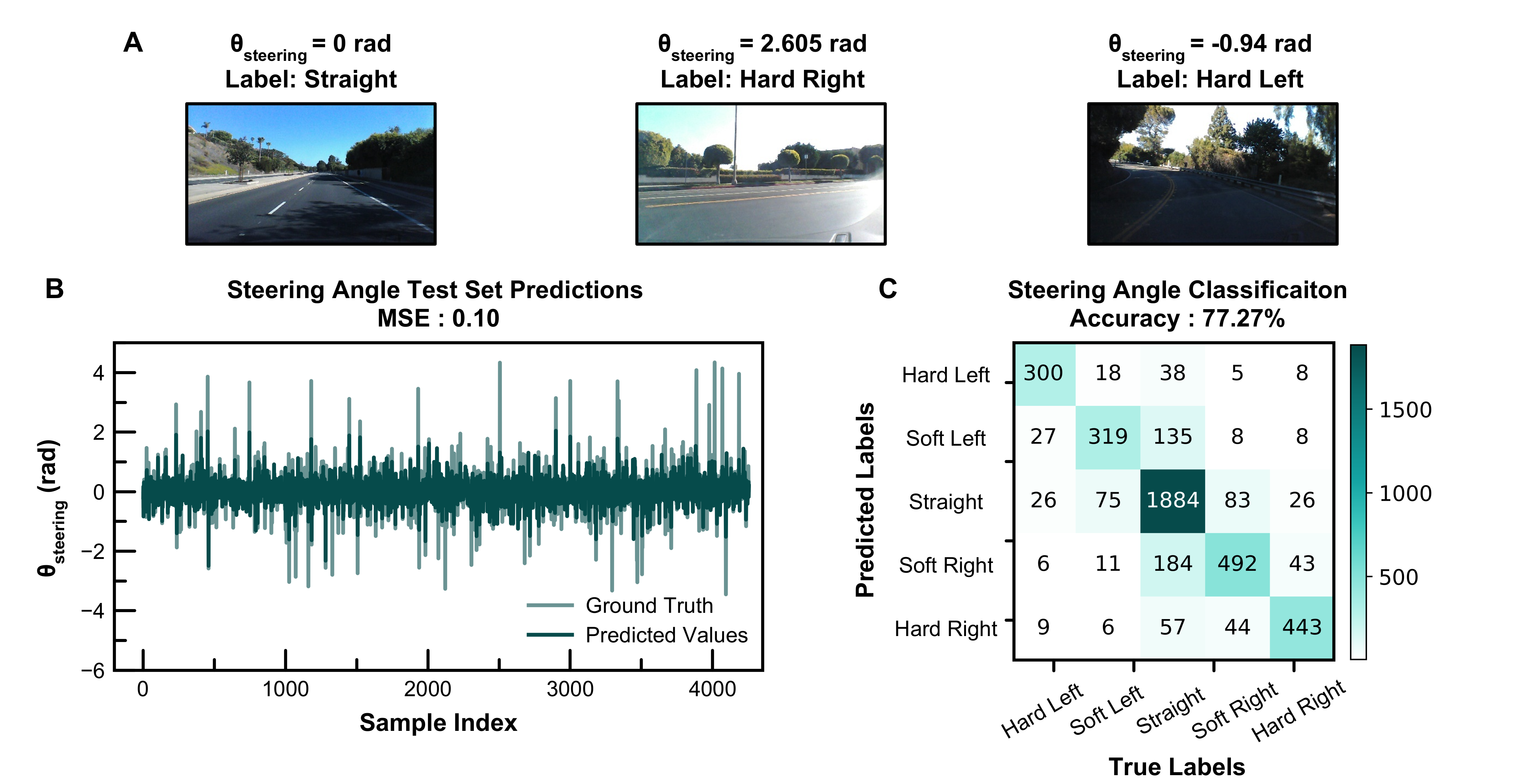}
\caption{\textbf{Steering angle prediction for autonomous driving.} (\textbf{A}) Representative dashboard-camera frames from different driving scenarios. (\textbf{B}) Continuous steering-angle regression showing close alignment between predicted and ground-truth steering commands. (\textbf{C}) Confusion matrix for five-class steering direction classification.}
\label{fig:driving}
\end{figure}

To evaluate real-world applicability, we test the system on a self-driving dataset containing dashboard-camera video paired with continuous steering-angle measurements \cite{chen_sullychendriving-datasets_2025}. This task requires the system to extract temporal context from dynamic road scenes and predict control signals based on visual flow and scene structure. Frame triplets from the driving videos are encoded and processed through the optical recurrent system. The resulting speckle features are used for both continuous regression of steering angles and five-class classification (hard left, soft left, straight, soft right, hard right). Figure \ref{fig:driving} illustrates the results of machine learning tasks on driving dataset.  The optical embeddings produce smooth steering predictions that closely track ground-truth control signals, demonstrating that passive recurrence captures the relevant spatiotemporal features for driving behavior modeling.

\subsection*{Surgical skill assessment}

Finally, we apply the optical processor to the JIGSAWS surgical skill dataset \cite{gao_jhu-isi_nodate}, which contains video recordings of robotic-assisted surgical tasks performd by operators with different skill levels. Skill assessment requires detecting subtle differences in motion smoothness, trajectory regularity, and temporal consistency, challenging features that depend on fine-grained spatiotemporal patterns. Videos are processed identically to previous experiments, with frame triplets encoded and accumulated speckle patterns serving as skill representations. Figure \ref{fig:driving} illustrates the results of machine learning tasks on surgical skill assessment dataset. The optical reservoir successfully classifies overall skill levels as well as specific performance metrics including respect for tissue and output quality. 

\begin{figure}[H]
\centering
\includegraphics[width=\textwidth]{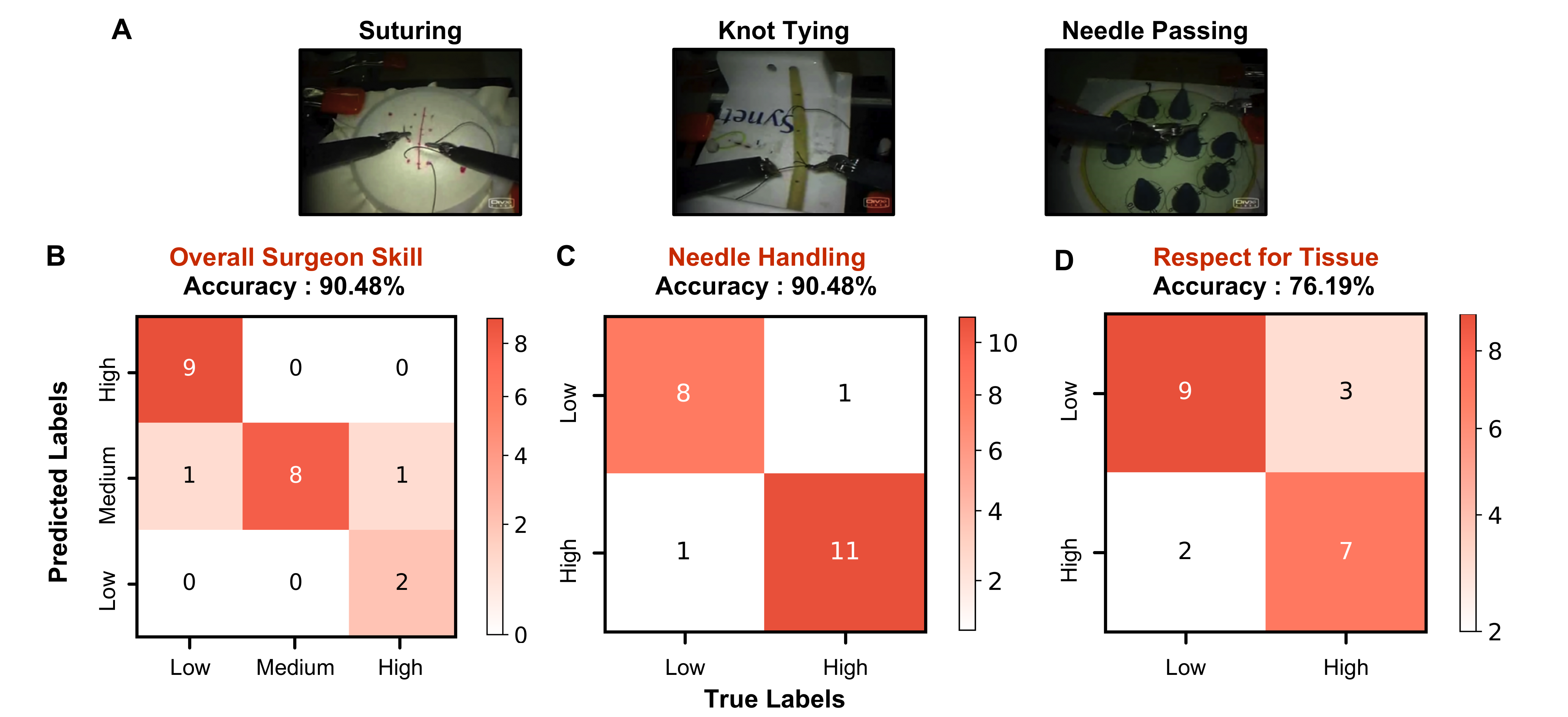}
\caption{\textbf{Surgical skill recognition from video.} (\textbf{A}) Representative frames from surgical task videos with performance metrics. (\textbf{B}) Confusion matrix of overall skill classification. (\textbf{C}) Confusion matrix of respect-for-tissue metric. (\textbf{D}) Confusion matrix of output-quality classification.}
\label{fig:surgery}
\end{figure}

\section{Discussion}

This work demonstrates that spatiotemporal dynamics in multimode optical fibers can serve as a physical computer for recurrent learning. By exploiting nonlinear propagation, modal mixing, and passive fiber-loop recurrence, the system generates high-dimensional optical states that encode both instantaneous inputs and a fading history without relying on digital memory or trained recurrent parameters. Such behavior aligns with recent studies on nonlinear multimode photonics and spatiotemporal mode interactions in multimode fibers \cite{wright2015controllable,wright2022nonlinear}. In contrast to photonic computing architectures that rely on interferometric stability, active modulation, or electronic feedback, the proposed approach performs sequence processing through intrinsic physical dynamics in a compact and passive platform.

A central implication of this work is that recurrence need not be implemented algorithmically or optimized through parameter learning. In conventional recurrent neural networks, temporal dependencies are encoded in trained weights and updated via backpropagation through time. Here, recurrence instead emerges as a physical process governed by beam propagation and delayed interference, where memory arises naturally from the circulation of optical fields. This perspective is consistent with delay-based and dynamical photonic reservoir systems \cite{Appeltant2011Reservoir,Larger2017PhotonicReservoir}, but extends these ideas to multimode spatiotemporal dynamics.

Across diverse tasks—including chaotic time-series forecasting, human action recognition, steering angle prediction, and surgical skill assessment—the same physical recurrent operator produces task-selective optical representations without any modification to the optical architecture. This universality indicates that task specificity arises from the interaction between input encoding and the intrinsic dynamics of the multimode fiber rather than hardware reconfiguration, in agreement with recent observations of multimode photonic neural networks operating near spatiotemporal chaos \cite{Kesgin2025ChaosMMF}. Notably, the experiments reveal that multiple aspects of spatiotemporal information are encoded simultaneously within the optical state: motion dynamics, temporal ordering, contextual background, and fine-grained execution differences are all reflected in the resulting speckle patterns. Task-relevant information can therefore be extracted by training only the readout layer, highlighting the expressive and multiplexed nature of multimode spatiotemporal dynamics. Beyond these scientific implications, the passive physical realization of recurrence also provides important practical advantages for scalable learning systems.

In practice, this passive architecture offers several advantages for scalable learning systems. Although the present free-space and fiber-based implementation requires careful optical alignment and calibration, the absence of active elements in the recurrent loop reduces sensitivity to timing jitter, thermal drift, and feedback instability, enabling stable operation over extended durations. Because recurrence is implemented through optical propagation rather than electronic state updates, the energy cost per recurrent update is minimal and largely independent of sequence length. These characteristics complement emerging photonic accelerators, optical tensor cores, and unconventional computing platforms such as coherent Ising machines \cite{Pierangeli2019Ising,mcmahon2016fully}.

A number of limitations remain. The current implementation relies on free-space coupling and precise alignment, which may limit robustness outside controlled laboratory environments. In addition, the temporal memory of the system is governed by passive attenuation and loop delay rather than trainable gating mechanisms such as those used in LSTM or attention-based architectures. As a result, the accessible memory horizon is fixed by physical parameters including loop length, loss, and dispersion. While the present results demonstrate stable operation across multiple tasks, systematic design rules for optimizing multimode recurrent processors—analogous to architecture design in electronic neural networks—are still lacking. Addressing these challenges will require improved integration, programmable photonic elements, and deeper theoretical understanding of multimode spatiotemporal dynamics.

Beyond the specific implementation demonstrated here, these findings point toward a broader class of learning systems in which physical dynamics perform the core recurrent computation while electronic processing is confined to readout and training. Recent works have explored related multimode-fiber and programmable optical neural network platforms \cite{Kesgin2025FiberD2NN,Carpinlioglu2025GeneticONN,Maula2025Supercontinuum}. Integrating multimode reservoirs with photonic integrated platforms and non-volatile photonic memory technologies may enable compact and scalable physical recurrent processors \cite{rios2015integrated,yildirim2024nonlinear}. 

Overall, recurrent harnessing multimode-fiber dynamics provide a promising route toward energy-efficient spatiotemporal learning. By embedding temporal processing directly in wave propagation, this work demonstrates that multimode optical systems can act as practical physical recurrent processors, with potential to expand as photonic integration and programmable optical technologies continue to mature.

\section{Materials and Methods}

\subsection*{Physical model of nonlinear propagation in a multimode fiber}
We describe light propagation in an optical fiber using the generalized (3+1)D Nonlinear Schrödinger Equation (NLSE)\cite{wright2017multimode,Agrawal2020NFO}. This equation governs the nonlinear operations of the arbitrary function $\mathcal{N}(A;\ell_{fiber})$ . The slowly varying envelope $A(x,y,t;z)$ inside a step-index multimode fiber obeys 
\begin{equation}
\begin{aligned}
\frac{\partial A}{\partial z} &=
\frac{i}{2k_0}\nabla^2_{\perp} A
-\beta_{1}\frac{\partial A}{\partial t}
-\frac{i\beta_{2}}{2}\frac{\partial^2 A}{\partial t^2}
+\frac{\beta_{3}}{6}\frac{\partial^3 A}{\partial t^3} 
+ i\frac{k_0}{2}(\frac{n_{core}^2}{n_{clad}^2}-1) A
+ i\,\mathbf{K}_{NL} .
\end{aligned}
\end{equation}

where $k_0 = 2\pi n_{core}/\lambda_c$ is the wavenumber, $n_{core} = 1.44$ and, $\beta_j$ are the dispersion coefficients. In this equation, $\mathbf{K}_{NL}$ is the nonlinear governing kernel that describes the nonlinear pulse propagation when $A$ is a single beam or is a superposition of two or three beams with different spatial mode profiles: 
\begin{equation}
\mathbf{K}_{NL} = 
\begin{cases}
\gamma \underbrace{|A|^2 A}_{SPM} & \text{for one beam} \\
\gamma (\underbrace{|A|^2}_{SPM} +  \underbrace{2|A_2|^2}_{XPM})A & \text{for two beams} \\
\gamma ( \underbrace{|A|^2}_{SPM} + \underbrace{2|A_2|^2 + 2|A_3|^2}_{XPM})A & \text{for three beams}
\end{cases}
\end{equation}

The nonlinear coefficient $\gamma$ represents the Kerr nonlinearity of the fiber. Since the wavelengths of the each pulse are the same in our setup we assume $\gamma = \gamma_1,\gamma_2,\gamma_3$  The nonlinear kernel accounts for SPM and XPM between co-propagating beams. As can be seen from the expression of the nonlinear kernel $\mathbf{K}_{NL}$, when $A$ is a superposition of multiple beams, SPM and XPM events occur simultaneously. The explicit derivations for case-specific NLSE are discussed in the Supplementary Note 4.

\subsection*{Experimental architecture}

The photonic recurrent processor is implemented using passive spatiotemporal dynamics in multimode optical fiber combined with fiber-delay encoding. All experiments use a pulsed laser source operating at 1030 nm with a pulse duration of approximately 400 ps and a repetition rate of 9 kHz.

The laser output is divided into three spatial channels using cascaded beam splitters (70/30 followed by 50/50), producing one beam with 35\% of the initial power and two beams with 30\% power. Mirrors directed the three beams onto the spatial light modulator (SLM). Free-space path lengths were equalized to remove relative delays prior to encoding and fiber coupling. Each beam is coupled into multimode fibers of different lengths using a $3\times1$ multimode fiber combiner. The pulse duration exceeds the modal dispersion accumulated over the 5–10\,m delay fibers, ensuring that the channels remain temporally separated at combiner section. All combiner input fibers are 105/125~$\mu$m step-index multimode fibers (NA = 0.22), while the output fiber is a 200/220~$\mu$m multimode fiber. The combined signal is coupled into a 50/50 fiber coupler forming a delayed feedback loop. Output speckle patterns are recorded using a camera positioned at the fiber output facet. Coupling efficiency and spectral measurements from the setup are outlined in Supplementary Notes 2-3 and Supplementary Figures 2-4.

\subsection*{Optical encoding and recurrent processing}

Three spatially multiplexed input channels are encoded on a reflective phase-only spatial light modulator (SLM) (see Supplementary Note 1). Three square regions of interest were used to encode temporally shifted frames of the input sequence corresponding to discrete times $n_0$, $n_0-1$, and $n_0-2$. Each triplet of frames formed a single SLM input.

The encoded wavefronts were coupled into multimode fibers using identical $2f$ relay optics. Fiber paths differed by fixed increments of $5~\mathrm{m}$ and $10~\mathrm{m}$ relative to the reference fiber, introducing deterministic arrival-time offsets
\[
\tau_i = \frac{n_{\mathrm{eff}}\Delta L_i}{c} = \frac{1.45 \cdot 5\,\text{m}}{299792458\,\text{m/s}} \approx 24\,\text{ns}
\]
The 5~m fiber increment matches the round-trip delay of the recurrent loop (see Supplementary Note 1). After the delay unit, a $50/50$ multimode fiber coupler split the signal such that part of the optical field recirculated inside the loop while the remainder was recorded as the output. Additional implementation details are provided in Supplementary Note 1 and Supplementary Figure 1.

\subsection*{Time-series forecasting protocol}

Forecasting experiments use the first half of Santa Fe laser time-series dataset consisting of 5,000 sequential samples. The first 3,500 samples are used for training in a teacher-forced configuration. At each time step, the ground-truth value is encoded on the SLM, propagated through the multimode-fiber reservoir, and the resulting speckle pattern is recorded to construct the training set for the readout layer.

After training, the system is evaluated in closed-loop forecasting mode using the remaining 1,500 samples. The final ground-truth training sample is used to initialize the reservoir state. For all subsequent time steps, the predicted output $\hat{y}_{n_0 + 1} \approx x_{n_0 + 1}$ was recursively re-encoded and used as the input to the reservoir without access to ground-truth values. This protocol evaluates the intrinsic temporal dynamics and stability of the physical recurrent processor under autonomous operation.

\subsection*{Video datasets}

The reservoir was evaluated on three video benchmarks: the KTH Human Actions dataset, the Sully Chen self-driving dataset, and the JIGSAWS surgical-skills dataset. Videos were converted into triplets of consecutive frames and encoded using the spatial–temporal embedding described above.

For all datasets, only frame triplets were provided to the optical system. Spatiotemporal mixing and recurrence occurred entirely within the optical reservoir. We take an incoherent sum of all reservoir states which are from each triplets. Each video sample was represented by a single speckle pattern used by the electronic readout for prediction.

\subsection*{Readout training}

Each sample is mapped to a single speckle pattern representing the optical response to three temporally adjacent frames. Task-specific predictions are obtained using linear regression or classification after L2 regularization. Models were trained and evaluated using dataset-specific protocols. For the KTH dataset, an 80/20 train–test split is used for general action classification and action-conditioned scene classification. For Scenario-1 action recognition and person identification under the running action, a 75/25 train–test split is used due to the small number of samples per class. For the self-driving, surgical skill, and Santa-Fe time series prediction datasets, an 80/20 train–test split was used. These choices affect only dataset splitting and evaluation and do not modify the optical processing pipeline. 


\clearpage 

\bibliography{reference_list} 
\bibliographystyle{sciencemag}

\paragraph*{Funding:}
This work is supported by the Scientific and Technological Research Council of Turkey (T\"{U}B\.{I}TAK) under grant number 123F171.

\paragraph*{Author contributions:}

\noindent Conceptualization: DE, BUK, UT

\noindent Methodology: DE, BUK, UT

\noindent Investigation: DE, BUK

\noindent Formal analysis: DE, BUK, UT

\noindent Writing---original draft: DE, BUK, UT

\noindent Writing---review \& editing: DE, BUK, UT

\noindent Funding acquisition: UT

\noindent Supervision: UT

\paragraph*{Competing interests:}
Authors declare that they have no competing interests. 

\paragraph*{Data and materials availability:}
Data used in the figures are publicly available at this online repository \cite{eslik_supplementary_2026}. Data generated during experiments may be obtained from the authors upon reasonable request. Datasets used in the study are publicly available through their respective sources: Santa Fe time-series data \cite{hubner_dimensions_1989,noauthor_mcnames_nodate}, KTH human actions \cite{schuldt_recognizing_2004}, Sully Chen self-driving dataset \cite{chen_sullychendriving-datasets_2025}, and JIGSAWS surgical skills \cite{gao_jhu-isi_nodate}. Code related to the results in this work may be obtained from the authors upon reasonable request.


\subsection*{Supplementary materials}
Supplementary Text\\
Figures S1 to S4\\

\end{document}